\documentclass[aps,prl,twocolumn,groupedaddress,superscriptaddress,preprintnumbers,amsmath,amssymb]{revtex4}
\usepackage{graphicx}
\usepackage[english]{babel}

\begin{document}
\title{ An all-silicon single-photon source by unconventional photon blockade }
    \author{Hugo Flayac}
    \email{hugo.flayac@epfl.ch}
     \affiliation{ Institute of Theoretical Physics, Ecole Polytechnique F\'{e}d\'{e}rale de Lausanne (EPFL), CH-1015 Lausanne, Switzerland }
    \author{Dario Gerace}
    \email{dario.gerace@unipv.it}
    \affiliation{Department of Physics, University of Pavia, I-27100 Pavia, Italy}
     \author{Vincenzo Savona}
     \email{vincenzo.savona@epfl.ch}
     \affiliation{ Institute of Theoretical Physics, Ecole Polytechnique F\'{e}d\'{e}rale de Lausanne (EPFL), CH-1015 Lausanne, Switzerland }
    \date{\today}
\begin{abstract}
The lack of suitable quantum emitters in silicon and silicon-based materials has prevented the realization of room temperature, compact, stable, and integrated sources of single photons in a scalable on-chip architecture, so far. Current approaches rely on exploiting the enhanced optical nonlinearity of silicon through light confinement or slow-light propagation, {and are based on parametric processes that typically require substantial input energy and spatial footprint to reach a reasonable output yield.}
Here we propose an alternative all-silicon device that employs a different paradigm, namely the interplay between quantum interference and the third-order intrinsic nonlinearity in a system of two coupled optical cavities. This {\em unconventional photon blockade} allows to produce antibunched radiation at extremely low input powers.  {We demonstrate a reliable protocol to operate this mechanism under pulsed optical excitation, as required for device applications, thus implementing a true single-photon source}. We finally propose a state-of-art implementation in a standard silicon-based photonic crystal integrated circuit {that outperforms existing parametric devices either in input power or footprint area}.
\end{abstract}
\maketitle

The last decade has witnessed a tremendous progress in silicon-on-insulator (SOI) technology for applications in photonic integrated computing and data processing \cite{Streshinsky2013}.
In parallel, integrated photonic circuits have become increasingly appealing to realize key tasks in quantum information and communication, thanks to their natural interfacing with long distance communication networks working in telecommunication band ($1.3 - 1.5$ $\mu$m wavelengths). Clearly, the combination of these two paradigms will likely allow to realize complex quantum operations on-chip that are far beyond what may be envisaged with table-top experiments, with significant and large scale impact on efficient and secure data processing and transmission~\cite{OBrien2013}.
Within this context, the generation of single photons plays a central role for the development of on-chip quantum photonic technologies~\cite{OBrien2009}. In particular, the recent advances in silicon-based quantum photonics~\cite{Politi2008,Politi2009,Spring2013a,Crespi2013} would strongly benefit from integrated single-photon sources on the same operating chip.

Single-photon sources on-demand can be realized with artificial two-level emitters, such as semiconductor quantum dots \cite{Michler2000,Pelton2002}, which have increasingly improved their radiative efficiency over the last few years \cite{Claudon2010,Reimer2012,He2013,Ates2013}. However, these single photon sources are typically based on III-V semiconductors, they work most efficiently at cryogenic temperatures, and integration with silicon-based nanophotonic circuits working in the telecommunication band \cite{Streshinsky2013} remains challenging. As a possible alternative, integrated single-photon sources in silicon-on-insulator (SOI) photonic circuits have been shown \cite{Davanco2012,Spring2013b,Thompson2013}, based on enhanced four-wave mixing induced by the silicon $\chi^{(3)}$ susceptibility and non-deterministic heralding. Even if the efficiency of such integrated sources can be improved by spatial multiplexing \cite{Collins2013}, compactness and scalability remain open issues.

An alternative route to single-photon generation relies on the photon blockade mechanism~\cite{Werner1999}, where a strong third-order nonlinearity in an optical resonator enables a shift of the resonant frequency by more than its linewidth when a single photon is already present. As a consequence, the device can absorb a photon only after the previous one has been emitted. However, this mechanism however requires a stronger optical nonlinearity than what is achieved in state-of-the-art SOI devices~\cite{Ferretti2012}.

Here, we build on the mechanism called {\em unconventional photon blockade} (UBP), recently advocated as a promising paradigm for single-photon generation~\cite{Liew2010,Bamba2011,Ferretti2013}. The UPB mechanism relies on quantum interference, and is therefore highly sensitive to an optical nonlinearity of small magnitude. At difference with the conventional blockade, it has been shown that UPB can also occur when the nonlinear frequency shift per photon is much smaller than the cavity linewidth, {which is usually the case also in silicon photonic crystal nanocavities}~\cite{Lai2014,Minkov2014,Dharanipathy2014}.
So far, such theoretical mechanism was only shown to work under continuous-wave (cw) excitation, which severely limits the usefulness of the proposed antibunched radiation as an actual single-photon source \cite{Bamba2011,Ferretti2013}.
{In the present paper we go beyond previous works on UPB by demonstrating a reliable protocol that allows to operate any such system under pulsed excitation. In fact, this can be technically achieved by a combination of excitation pulse tailoring and post-selective temporal filtering of the output stream to purify the statistics of the emitted radiation, similarly to what has been already demonstrated for quantum dot-based single photon sources \cite{Ates2013}. The latter achievement allows to overcome a previously believed limitation, and provides a scheme to devise a true single photon source out of a generator of antibunched radiation.
The efficiency of such a single photon source in realistic devices is analyzed in detail, which is shown to outperform the best heralded sources demonstrated so far in key figures of merit, especially operation power and spatial footprint.}

%
%
\begin{figure}[t]
\includegraphics[width=0.48\textwidth, clip = true]{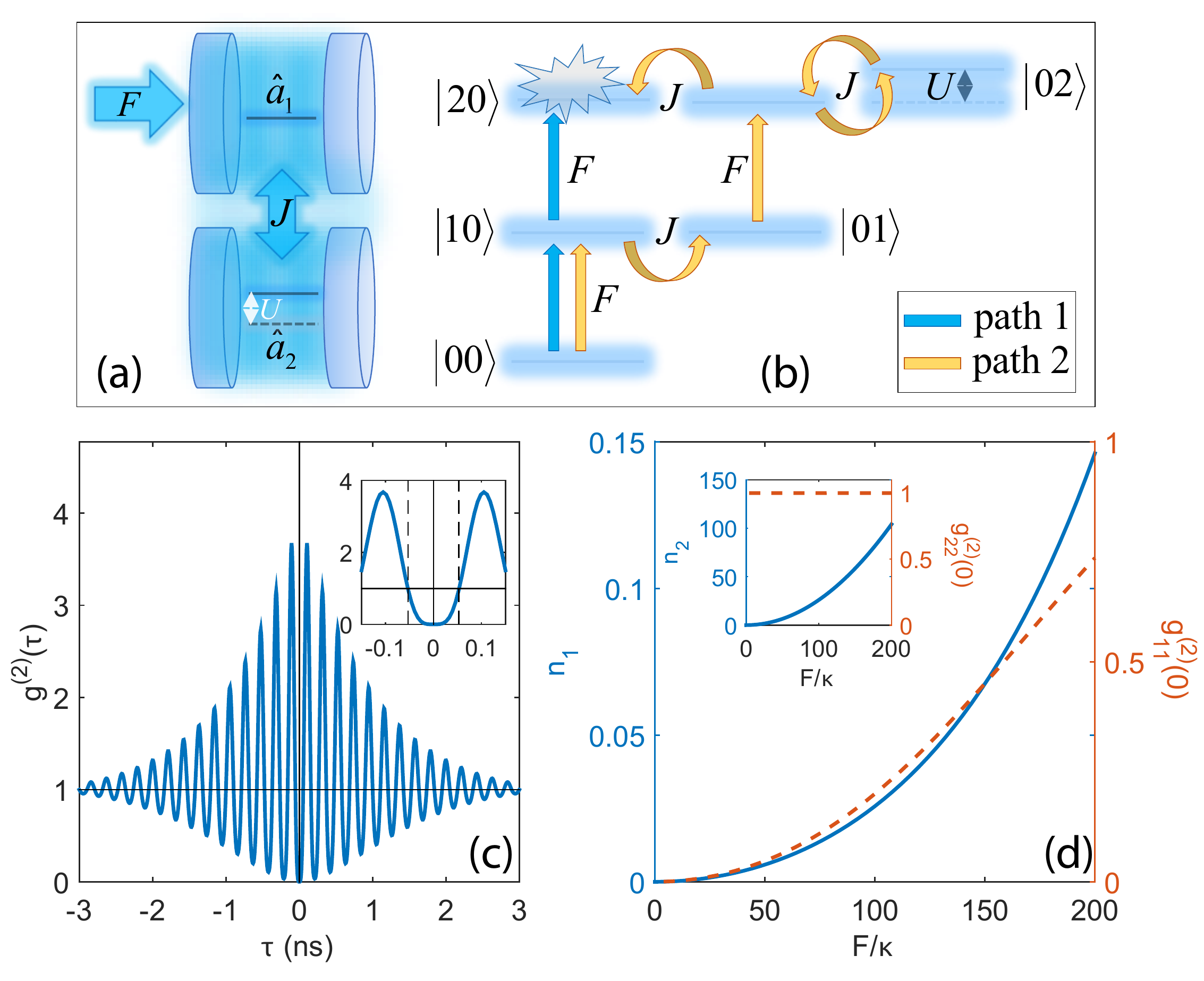}
\caption{\textbf{Unconventional photon blockade}.
(a) Schematic representation of an asymmetrically driven photonic molecule. Each cavity is characterized by a single resonant mode in the spectral region of interest, and only the first cavity is driven by an external coherent field.
(b) The corresponding ladder scheme of the lowest few energy levels, associated to photon occupation number states in the two cavities.
(c) The computed second-order correlation function of the quantum field in the first cavity, $g^{(2)}(\tau)$, plotted as a function of time delay. The quantity was computed under the assumption of a cw driving field. Inset: a detail of the photon antibunching region close to zero delay.
(d) Average photon occupation of the first cavity ($n_1$, full line), and corresponding value of $g^{(2)}(0)$ (dashed line), computed as a function of the cw driving field amplitude.
Inset: Same quantities plotted for the second cavity.
   }
\label{fig1}
\end{figure}

\vspace{0.4cm}
{\bf Results}
\vspace{0.4cm}

Following Refs. \onlinecite{Liew2010,Bamba2011}, we consider UPB in a system of two tunnel-coupled cavities, i.e. a {\em photonic molecule}, as sketched in Fig.~\ref{fig1}a. The quantum model of UPB has been thoroughly characterized \cite{Ferretti2013}, and it is briefly summarized in the Methods section. The relevant physical parameters are the tunnel coupling rate between the two cavities ($J/ \hbar$), the driving rate on the first cavity ($F/\hbar$), the driving frequency ($\omega_L$), the effective photon-photon interaction energy in each cavity ($U_j$, $j=1,2$), which originates from the intrinsic material $\chi^{(3)}$ \cite{Ferretti2012}, and the cavities loss rates $\kappa_j$. Detrimental pure-dephasing processes are known to be negligible if the overall dephasing rate is much smaller than $U_j/\hbar$ \cite{Ferretti2013}. For a photonic crystal cavity in silicon, this condition is largely fulfilled \cite{Flayac2014}. Finally, the model can be generalized to include input and output quantum channels \cite{Flayac2013}. We will assume $\omega_j =\omega_c$, $U_j = U$ and $\kappa_j = \kappa$ in the following, without loss of generality.

A scheme of the lowest 6 levels on the basis of photon-number states, $|n_1 \, n_2 \rangle$, is given in Fig.~\ref{fig1}b. The different excitation pathways leading from the initial ground state to the state $|2 \, 0 \rangle$ -- corresponding to two-photon occupation of the first (driven) cavity -- are highlighted. The UPB is essentially based on suppression of such double occupation by a careful tuning of the model parameters, leading to destructive quantum interference between the two alternative pathways.
The optimal UPB conditions \cite{Bamba2011} are given by $J_{\mathrm{opt}} / \hbar\kappa \simeq [(2/3\sqrt{3})\hbar\kappa/U]^{1/2}$, and $\Delta_{\mathrm{opt}}=(\omega_c - \omega_L)=-\kappa/2\sqrt{3}$, and will be assumed to hold in what follows.

We consider the UPB mechanism in a SOI nanophotonic platform, where the cavity-field confinement in a diffraction-limited mode volume, $V \sim (\lambda/n)^3$, enhances the effective photon-photon interaction, $U$. A realistic order of magnitude estimate in a crystalline silicon photonic crystal nanocavity leads to $U\simeq 10^{-3}$ $\mu$eV (see also Supplementary Information) \cite{Ferretti2013}. Assuming a quality factor $Q \simeq 8\times 10^5$ -- now routinely achieved at telecom wavelengths (i.e., $\hbar\omega_c \sim 0.8$ eV) \cite{Lai2014,Notomi2010,Sekoguchi2014} -- we set $\hbar\kappa \simeq 1$ $\mu$eV, and hence $U/ \hbar\kappa = 0.001$.
To fulfill the optimal UPB conditions, the remaining parameters take values $\Delta = - 0.29 \kappa$ and $J = 19.6 \hbar\kappa$, respectively.

The steady state results under cw driving are summarized in Fig.~\ref{fig1}c-f. The time-dependent normalized second-order correlation function, $g^{(2)}(\tau)$ (see Methods section) is considered as the reference figure of merit for single-photon blockade~\cite{Birnbaum2005,Faraon2008,Reinhard2012} and plotted in Fig.~\ref{fig1}c.
A strong antibunching is present over a time-delay window $\tau < 100$ ps. At longer delays, strong oscillations are present on the timescale $h/J$, arising from the interferential nature of the UPB mechanism \cite{Bamba2011}. The average photon occupations in the two cavities, $n_i=\langle\hat{a}_i^{\dagger}\hat{a}_i\rangle$, and the corresponding zero-delay correlation, $g^{(2)}(0)$, are displayed as a function of the driving field amplitude, $F$, in Fig.~\ref{fig1}d: UPB occurs at low average occupation of the driven cavity, while the occupation of the non-detected cavity is much larger (see inset). This figure of merit is relevant to determine the maximal single-photon emission rate that can be achieved in such a device under cw pumping, given by $R_{\mathrm{em}}= n_1 \kappa/2\pi$. As an example, for $n_1 \simeq 0.05$ (corresponding to $F/ \hbar \kappa \sim 30$ and $g^{(2)}(0) < 0.5$), $R_{\mathrm{em}} > 10$ MHz can be expected, with an input power as low as $P_{\mathrm{in}}=\hbar\omega_c F/h\simeq 1$ nW.
{In fact, the optimal UPB relation between $U$ and $J$ leads to a condition (without numerical pre-factors, for convenience) $J_{\mathrm{opt}}\propto \sqrt{V}/Q^2$; this means that the required input-power in UPB scales down roughly as $1/Q^2$, i.e. the larger the cavity Q, the smaller $J_{\mathrm{opt}}$ can be to have antibunching by keeping the average number of photons in the first cavity less than 0.1 (according to Fig.~\ref{fig1}d). The same figure of merit simultaneously allows to increase the antibunching time window, scaling as $1/J_{\mathrm{opt}}$ (see Fig.~\ref{fig1}c). }

%
%
\begin{figure}[t]
\includegraphics[width=0.48\textwidth, clip = true]{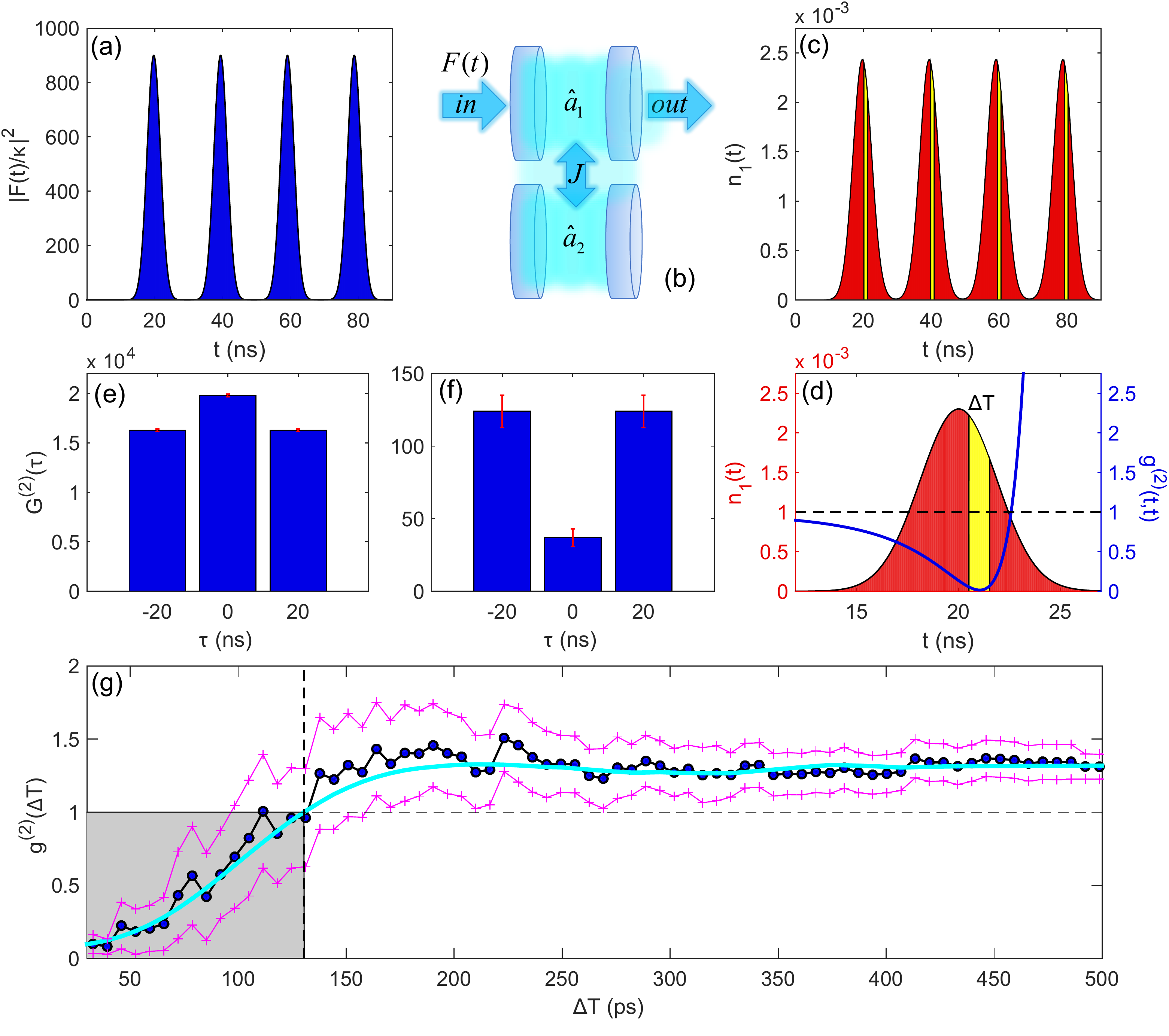}
\caption{ \textbf{Unconventional photon blockade under pulsed excitation}.
(a) Pulse sequence driving the first cavity, as schematically represented in (b). Pulse duration is set to $\sigma_t=4$ ns at $50$ MHz repetition rate.
(c) Corresponding cavity output, i.e. average population in the driven cavity as a function of time. The filtering window is schematically superimposed within each pulse (yellow areas).
(d) Detail of a single pulse from the output sequence in panel (c). The blue curve shows the equal time second-order correlation function, $g^{(2)}(t,t)$ (scale on the right).
(e) Un-normalized and un-filtered second-order correlation over the whole pulse sequence (see Supplementary Information).
(f)  Same, after time-filtering the pulses with $\Delta T=75$ ps.
(g) Full Montecarlo wave function simulation of the time-filtered second-order correlation function under pulsed excitation, as a function of the width $\Delta T$ of the filtering window (disks), and relative error span (crosses). The cyan line is the two-time second-order correlation calculated by solving the quantum master equation (see Supplementary Information). The model parameters assumed in these simulations are the same as in Fig.~\ref{fig1}.
  }
\label{fig2}
\end{figure}

Single-photon sources on demand require the emission of single-photon pulses at deterministic times.
However, in the UPB mechanism the emitted light is sub-Poissonian only within a time delay shorter than $h/J$, as shown in Fig.~\ref{fig1}c. For short input pulses, the outgoing pulses will last at least as long as the cavity lifetime. Therefore, the condition $J\gg\hbar\kappa$ would apparently prevent the device from operating in a pulsed regime \cite{Bamba2011}.

Here we show {for the first time} that UPB under pulsed excitation is possible by exploiting temporal filtering of the output signal. In Fig.~\ref{fig2}, the results of a numerical experiment are reported for the UPB device considered in the previous section, where a train of gaussian pulses drives the first cavity (Fig.~\ref{fig2}a-c). A sequence of outgoing pulses from cavity 1 is modeled by solving the quantum master equation (see Supplementary Information for details), and shown in Fig.~\ref{fig2}c.
Focusing on a single outgoing pulse, the equal-time second-order correlation is plotted in Fig.~\ref{fig2}d (blue curve), where a well-defined time window clearly exists -- within the pulse emitted from a UPB device -- during which light is antibunched over a time delay shorter than $h /J$.

As a consequence, pulsed operation can be achieved by gating the outgoing pulses in time, in order to retain only a timeframe of duration $< h /J$. In practice, this could be achieved with an integrated all-optical switch triggered by the input pulse, as it was already shown experimentally \cite{Ates2013}. The second-order correlation function under pulsed excitation (see Supplementary Information) is shown in Figs.~\ref{fig2}e-g.
The histograms in Figs.~\ref{fig2}e-f show the un-normalized correlation signal, $G^{(2)}(\tau)$, integrated over the whole pulse sequence in (Fig.~\ref{fig2}e) and in the presence of filtering with a time window $\Delta T=75$ ps (Fig.~\ref{fig2}f), respectively. They reveal a strong reduction of the two photon counts within a pulse after filtering, which is a key result of this paper. Figure~\ref{fig2}g displays the dependence of the filtered second-order correlation versus the filtering time window, $\Delta T$.
The Montecarlo data (blue disks), directly obtained from the photon count statistics (see Supplementary Information), are reproduced by a master equation treatment (cyan curve), which confirms the reliability of this result. Photon antibunching (gray area) is achieved below $\Delta T=130$ ps, while the single photon regime -- requiring the condition $g^{(2)}(\Delta T)<0.5$ -- is obtained for $\Delta T < 90$ ps.
When assuming $5\times 10^7$ pulses per second and a peak value $F\sim 150\hbar\kappa$, after the temporal filtering the Montecarlo data indicate a single-photon yield at a rate of about 0.45 MHz.
Under these conditions, the driven cavity reaches a peak value of the average photon occupation $n_1\sim 0.075$, close to the largest occupancy for which UPB is expected according to Fig.~\ref{fig1}c [i.e., $g^{(2)}(0) < 0.5$]. Remarkably, this peak value of $F(t)$ implies an intracavity energy of less than $10^{-2}$ fJ per pulse. {This corresponds to an input energy that can be quantified as $\sim 0.5$ fJ per pulse, according to typical excitation schemes of photonic crystal integrated circuits \cite{Dharanipathy2014}.
We notice that this is extremely low when compared to the state-of-art parametric sources demonstrated so far in integrated silicon-based platforms and based on four-wave mixing and heralding  (typical input powers in the 100 mW range), for a comparable output rate in the few 100 kHz range \cite{Davanco2012,Spring2013b,Thompson2013}. }

%
%
\begin{figure}[t]
\includegraphics[width=0.48\textwidth, clip = true]{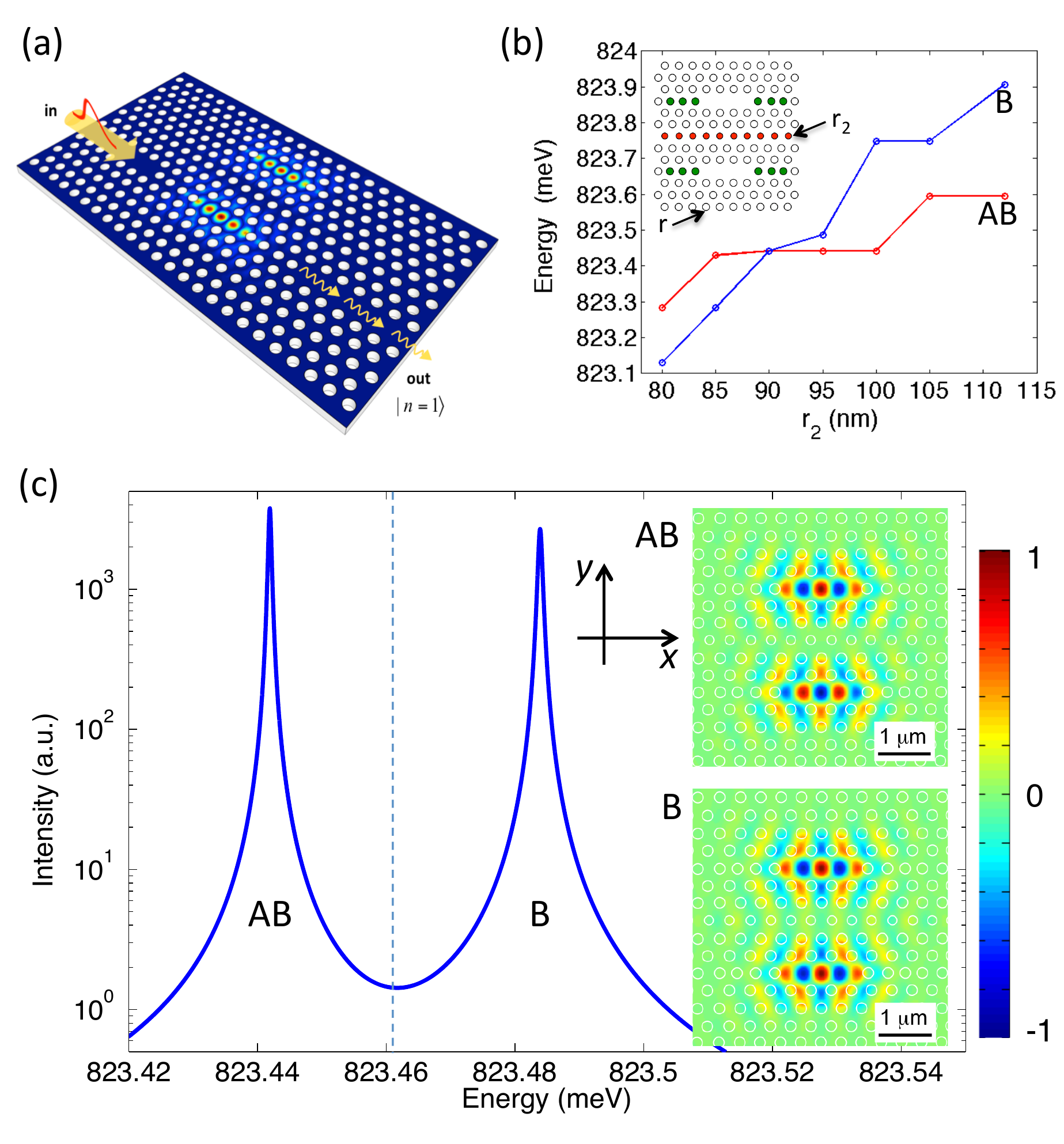}
\caption{ \textbf{Realization of a SOI integrated single-photon source}.
  (a)  Artistic view of an integrated SOI photonic crystal chip realizing input/output channels and UPB through a photonic crystal molecule.
  (b)  Fine-tuning of the normal mode splitting of a L3 photonic crystal molecule (see text) through variation of the radius ($r_2$) in the middle row of holes (red-highlighted in the inset). The holes highlighted in green are shifted off from the cavities center to optimize the Q-factor \cite{Lai2014,Minkov2014}. The hole radius of the surrounding photonic crystal lattice is $r=112$ nm. All the design parameters are given in detail in the Supplementary Information associated to this manuscript.
  (c)  Spectrum of the photonic crystal molecule designed to have the parameters corresponding to the results shown in Figs.~\ref{fig1}-\ref{fig2}. The two normal mode resonances are indicated as bonding (B) and antibonding (AB), respectively. The exciting laser frequency at the optimal UPB condition is schematically indicated (vertical dashed line).
Inset: $E_y$ component plotted for the two normal modes, superimposed to the photonic crystal design showing the footprint area of the coupled cavities device. The reference directions $x$ and $y$  are explicitly indicated.
 }
\label{fig3}
\end{figure}

A feasible realization is hereby proposed in an integrated SOI photonic crystal platform. These structures benefit from a remarkably advanced and well established fabrication technology, with nanocavities having recently achieved $Q/V$-values well above the UPB requirements~\cite{Lai2014,Dharanipathy2014,Notomi2010,Sekoguchi2014}.
As a schematic example, we show in Fig.~\ref{fig3}a a double-cavity device in a photonic crystal circuit. This configuration allows to selectively drive one cavity from the input waveguide channel, and simultaneously collect part of the light emitted from the same cavity into the output waveguide (the remaining part being emitted through out-of-plane losses).
As an elementary building block, we consider a L3 photonic crystal cavity in a thin silicon membrane, designed for operation at the preferred telecom wavelength, $\lambda = 1.5$ $\mu$m ($\sim 0.825$ eV). The cavity consists of three missing air holes in a triangular air-hole lattice. This cavity was recently optimized to show a measured quality factor regularly exceeding one million~\cite{Lai2014,Minkov2014}. We use here a L3 cavity design with theoretical unloaded (i.e., valid for the isolated cavity) $Q\sim 1.25 \times 10^6$ (see Supplementary Information for details on the structure parameters, such as hole radius and lattice constant). When coupling to the access waveguides, the loaded Q-factor can be engineered in the range $Q\sim 8 \times 10^5$, as we have verified by 3D finite-difference time-domain simulations (3D-FDTD, not shown).
From the calculated mode profile, the effective nonlinearity for this device is estimated (see Supplementary Information) in the range $U\simeq 0.8\times 10^{-3}$ $\mu$eV, close to what was assumed in the model calculations above.

The photonic crystal molecule can be obtained by vertically aligning two L3 cavities, separated by 5 rows of holes (i.e., $3\sqrt{3}a$ center-to-center)~\cite{Chalcraft2011}. The hole radius in the central row, $r_2$, can be varied to fine tune the normal modes splitting at the desired value~\cite{Haddadi2014}, i.e. $\Delta = 2 J$ according to Eq.~\ref{eq:ham} in Methods. In Fig.~\ref{fig3}b we show such a simulated fine tuning by 3D-FDTD calculations. For $r_2 \sim 95$ nm, the normal mode splitting between the two resonances, identified as bonding (B) and antibonding (AB) according the the spatial profile of the $E_y$ component, is calculated as $\Delta \simeq 44$ $\mu$eV, i.e. remarkably close to the condition $J/ \hbar\kappa = 19.6$ assumed in the previous calculations, when we consider the loaded value $\hbar\kappa \simeq 1$ $\mu$eV.
We notice that similar values and dynamic control of the normal mode splitting have been already shown experimentally in SOI photonic crystal platforms operating in a very similar wavelength range~\cite{Sato2012}.
The spectrum for such an optimal structure is shown in Fig.~\ref{fig3}c. The two resonances have unbalanced Q-factors of $\sim 1.1 \times 10^6$ (AB) and $\sim 1.3 \times 10^6$ (B), respectively, which can also be exploited to enhance the degree of antibunching in UPB \cite{Ferretti2013}.

{Before concluding, we discuss how to circumvent the most relevant and potentially detrimental effects for the realization of UPB in a SOI platform.
First of all, two-photon absorption (TPA), related to the imaginary part of the silicon $\chi^{(3)}$, is also enhanced by confinement in the L3 cavities. However, a quantitative estimate of this contribution has been given in Ref. \onlinecite{Ferretti2012}, by which we can infer a TPA loss rate that is on the order of $\kappa_{TPA}/ \kappa < 10^{-2}$ for the present case, also considering the low input powers necessary to achieve UPB.
Second, thermal effects can give rise to pure dephasing of the cavity resonances, which depends on optomechanical coupling with the background phonons. For the L3-type silicon photonic crystal cavities considered here, this contribution has been quantitatively estimated and shown to be negligible even at room temperature (i.e., a pure dephasing rate
$\gamma^{\ast}/ \kappa \sim 10^{-7}$) \cite{Flayac2014}.
Finally, unavoidable fabrication imperfections should be corrected by device post-fabrication processing. In particular, fine and selective cavity tuning has been already shown for photonic crystal cavities with different techniques, even in the presence of very large Q-factors \cite{Sato2012,Caselli2013,Waks2013apl}.}

\vspace{0.4cm}
{\bf Discussion}
\vspace{0.4cm}

We have theoretically shown that an integrated nanophotonic platform based on CMOS-compatible SOI technology can be engineered to achieve single photon emission by an unconventional photon blockade mechanism. Besides opening the way to the first experimental demonstration of UPB, our results show that such unconventional mechanism allows for pulsed excitation, which represents a key ingredient for a useful source where each pulse potentially triggers emission of a single photon.

{Such an alternative single-photon source could be characterized by a very low input power operation, i.e. comparable to standard single-photon devices based on cavity QED but much lower than typical integrated single-photon sources based on enhanced four-wave mixing and heralding.}
Moreover, this is achieved by an unprecedented small footprint area, significantly smaller than recently realized heralded sources in integrated SOI chips. In fact, notice that the footprint of this prospected device is essentially given by the spatial extension of the photonic crystal molecule and the necessary lattice around it. For the structure simulated in Fig.~\ref{fig3}, we estimate a minimal footprint area on the order of a few $\mu$m$^2$ (see, e.g., the inset in Fig.~\ref{fig3}c). In practice, this is significantly smaller than current heralded sources fabricated with the same SOI technology and based on coupled resonator optical waveguides \cite{Davanco2012} or spatially multiplexed photonic crystal waveguides \cite{Collins2013}.
We also stress the generality of the proposed scheme, which could be extended to other types of nonlinearities \cite{Gerace2014}, and could eventually lead to the realization of novel quantum devices \cite{Gerace2009,Dudu2014} for applications in integrated quantum metrology and logic.

In summary, {by combining an extremely low input power, a small footprint area, and no quantum emitter required for single-photon generation, such a device might have significant impact on the development of integrated silicon quantum photonics, by introducing a new concept in the generation of pure quantum states of light at arbitrary wavelengths (e.g., in the telecom band), that is fully compatible with current semiconductor technology, working at room temperature, and a viable alternative to single-photon nonlinear devices based on cavity-QED with artificial atoms or single atomic-like emitters that are presently lacking in SOI integrated platforms.}

\begin{small}
\vspace{0.4cm}
{\bf Methods}
\\
\textit{Model Hamiltonian}.
The second quantized hamiltonian of the driven nonlinear photonic molecule is expressed (to leading linear and nonlinear orders) as \cite{Liew2010,Bamba2011,Ferretti2013}
\begin{eqnarray}\label{eq:ham}
\hat{{\cal{H}}}_s &=& \sum_{i=1,2}[\hbar\omega_{i} \hat{a}_{i}^{\dag}\hat{a}_{i}+
U_i \hat{a}_{i}^{\dag}\hat{a}_{i}^{\dag}\hat{a}_{i}\hat{a}_{i}] + J(\hat{a}_{1}^{\dag}\hat{a}_{2}+\hat a_{2}^{\dag}\hat{a}_{1}) \nonumber \\
&+& F(t) e^{-i\omega_L t} \hat{a}_{1}^{\dagger}+F^{\ast}(t) e^{i\omega_L t}\hat{a}_{1}
 \, .
\end{eqnarray}
The first terms in Eq. \ref{eq:ham} describe two harmonic oscillators, $J/ \hbar$ is the tunnel coupling rate between the two resonators, $F(t)/\hbar$ is the coherent pumping rate on the first cavity at the laser frequency $\omega_L$, and the photon-photon interaction energy in each resonator is related to the material $\chi^{(3)}$ \cite{Ferretti2012,Ferretti2013}.
A description of this quantity and an estimation for the photonic crystal cavities considered here are given in the Supplementary Information.
Starting from this Hamiltonian, the various time-dependent photon correlation functions for light collected after cavity 1, generally defined as
\begin{equation}\label{eq:g2tau}
g^{(2)}(t,t')=
\frac{\langle\hat{a}_1^{\dagger}(t)\hat{a}_1^{\dagger}(t')\hat{a}_1(t')\hat{a}_1(t)\rangle}
{\langle\hat{a}_1^{\dagger}(t)\hat{a}_1(t)\rangle \langle\hat{a}_1^{\dagger}(t')\hat{a}_1(t')\rangle} \, ,
\end{equation}
were numerically simulated by using both the Montecarlo wave function method and by directly solving the master equation for the density matrix (see details in Supplementary Information).
In both cases, the numerical solutions were computed on a truncated Hilbert space of dimensions
 $(N_1\times N_2)^2$, where $N_1=4$ and $N_2=18$ are the maximum photon occupations allowed in cavities 1 and 2, respectively. While the master equation results were obtained from a modern workstation embedding 16 Gb of RAM, the Montecarlo data were produced by 10 nodes of 16 cores and 32 Gb RAM memory each, run on a high-end cluster for a few weeks of continuous computational time.


\vspace{0.4cm}
\textbf{Acknowledgments}\\
Several useful discussions with A. Badolato, I. Carusotto, S. Ferretti, M. Galli, A. Imamo\v{g}lu, T. F. Krauss, T. C. H. Liew, L. O'Faolain, K. Srinivasan, H. E. T\"{u}reci, and J. P. Vasco, are gratefully acknowledged.
The authors acknowledge the Swiss National Science Foundation for support through the International Short Visits program, project number IZK0Z2-150900.
D.G. acknowledges partial financial support from the Italian Ministry of University and Research through Fondo Investimenti per Ricerca di Base (FIRB)
``Futuro in Ricerca'' project RBFR12RPD1.



\end{small}


\pagebreak
\begin{widetext}
\begin{center}
\textbf{\large Supplementary Information to \\ ``An all-silicon single-photon source by unconventional photon blockade''}
\end{center}
We detail here the Montecarlo and master equation treatments used to produce the results of figures 1 and 2 in the main text. We also give details on the photonic crystal cavities design, and the corresponding estimation of the single-photon nonlinearity reported in the manuscript.
\end{widetext}
\setcounter{equation}{0}
\setcounter{figure}{0}

\section{Montecarlo wave function method}

The statistics of the photons emitted by the system under pulsed excitation were first addressed by performing quantum Montecarlo simulations \cite{Dum1992}. This method not only allows to work with larger truncated Hilbert spaces but also provides direct access to individual photon counts, thus embodying the closest theoretical simulation of an actual Hanbury Brown-Twiss experiment. In brief, the algorithm is based on the stochastic evolution of the system wave function through the Schr\"{o}dinger equation
\begin{equation}\label{Sch}
\hat{{\cal{H}}}\left| \psi  \right\rangle  = i\hbar \frac{\partial }{{\partial t}}\left| \psi  \right\rangle  \, ,
\end{equation}
written for the non-Hermitian effective Hamiltonian
\begin{equation}\label{Heff}
\hat{{\cal{H}}} = {\hat{{\cal{H}}}_{s}} - \frac{{i\hbar }}{2}\sum\limits_j {\kappa_j\hat a_j^\dag {{\hat a}_j}} \, ,
\end{equation}

The non Hermitian part of \ref{Heff} results in a decay of the norm $\left\langle {\psi }\mathrel{\left | {\vphantom {\psi  \psi }}\right. \kern-\nulldelimiterspace}{\psi } \right\rangle$. During the evolution of Eq.~\ref{Sch}, random numbers $0<r<1$ are drawn and the condition $\left\langle {\psi }\mathrel{\left | {\vphantom {\psi  \psi }}\right. \kern-\nulldelimiterspace}{\psi } \right\rangle \le r$ decides for the action of a jump operator, $\hat{a}_j\left| \psi  \right\rangle$, corresponding to the measurement of a photon. The proper quantum jump operator is chosen such that $j$ is the smallest integer satisfying $\sum\nolimits_j P_j  \ge r$, where $P_j$ are the probabilities for the mode $j$ to emit a photon at a given time. Each evolution of Eq.~\ref{Sch} produces a stochastic quantum trajectory associated with the state ${{{\left| \psi{\left( t \right)}  \right\rangle }_j}}$, and the procedure can be repeated $N$ times to form an ensemble average of realizations in view of approximating the system density matrix as $\hat \rho \left( t \right) \mathop {\rm{ = }}\limits_{N \to \infty }  {\left| {\Psi \left( t \right)} \right\rangle}{\left\langle {\Psi \left( t \right)} \right|}$, where
\begin{equation}
\left| {\Psi \left( t \right)} \right\rangle  = \sum_{j = 1}^N {{{\left| {\psi \left( t \right)} \right\rangle }_j}}/N \, ,
\end{equation}
and its potential mixed nature. Any observable or correlation are obtained from $\langle {\hat O( t )}\rangle  = {\rm{Tr}}[ {\hat \rho ( t )\hat O} ]$. The full access to photon counts and emission times history allows to mimic the experimental detection scheme. The two-times second-order correlation function, $g^{(2)}(t_1,t_2)$, can be reconstructed from the statistics of photons delays analogously to a Hanbury Brown and Twiss (HBT) setup. Further details on the numerical procedure employed to obtain the results of Fig.~2 in the main text are given in the following.

\section{Two-time correlations under pulsed excitation}

In our Montecarlo simulations we have worked on the basis of trajectories containing single pulses, which makes the data analysis more flexible. We have tracked the quantum jumps performed by the driven cavity from which the photon antibunching is expected. We point out that in reality one should expect a weak mixing between both cavity fields to occur in the guiding channels. It can be accounted for within an input-output treatment. In such a case the laser detuning should be properly adapted according to the prescriptions of Ref. \onlinecite{Flayac2013}. To analyze delays within a given pulse we need to track the trajectories where \emph{at least} two quantum jumps occurred within $\Delta T$ and these events are obviously rare given the relatively small occupation of the cavity 1 as one can see from Fig.~2 of the main text. We have therefore performed a large campaign of massively parallelized simulations on an high end cluster based on $N = 1.5 \times10^{8}$ pulses from which we have recorded the whole quantum jump history. We considered pulses of duration 4 ns separated by 20 ns to avoid any overlap bringing some unwanted pulse to pulse correlations. Our simulation therefore covers not less than 1 seconds of recorded data.

To build the Monte-Carlo curve of Fig. 2g (blue disks), we have worked on quantum jumps that occurred in a time window of width $\Delta T=T_2-T_1=1.57$ ns centered on the $g^{(2)}(t,t)$ minimum (see yellow surface and blue curve of Fig. 2f) mimicking the temporal filtering. Inside this global window we have considered sub-windows of variable duration $\Delta t=t_2-t_1$ ranging from 6.5 ps to $\Delta T$. Each of these sub-windows was gradually displaced by $\Delta t$ within $\Delta T$, starting from the condition $t_1=T_1$ ($t_2=t_1+\Delta t$) and until $t_2=T_2$ ($t_1=t_2-\Delta t$) is fulfilled. For a given value of $\Delta t$, the un-normalized second order correlation $G^{(2)}(\Delta t)$ is obtained from the sum of photon pair counts recorded by slicing the time window, which increases the statistics by $N_w=\Delta T/\Delta t$. Therefore it allows to work with a number of counts that would correspond to $N \times N_w$ trajectories (pulses) reducing by $N_w$ the required computational time. Finally, the $g^{(2)}(\Delta t)$ is obtained by normalizing $G^{(2)}(\Delta t)$ to the number double counts expected from a Poissonian statistics. The errors (magenta curve) are computed from the square root of the number of counts averaged over the sliding windows. Obviously the error is inhomogeneous versus $\Delta t$, given that $N_w$ is variable, and it is small both in the regions of narrow and wide $\Delta t$ where respectively $N_w$ is large and number of counts is important. The previously described procedure is summarized in Fig.~S1 (see captions).

To build the histograms of Fig.~2d-e displaying the pulse-to-pulse statistics, we have performed a Montecarlo rearrangement of our single pulse trajectories to randomize their time ordering, as it would be obtained from many pulse trajectories or in an actual experimental situation.

\begin{figure}[h]
\includegraphics[width=0.5\textwidth,clip]{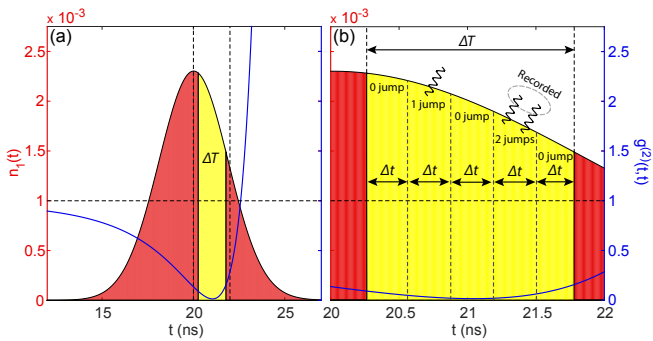}
\caption{(a) Average occupation of cavity 1 (red) and the equal time second order correlation function (blue line). The yellow region highlights the global time window $\Delta T$ from which the photon counts are extracted. (b) Zoom in between the vertical dashed lines of panel (a) showing an illustrative set of sub-windows for a given value of $\Delta t$. The wavy lines illustrate and example of quantum jump series for a given trajectory. Only the 2 jumps event are kept for the $G^{(2)}(\Delta t)$ statistics. 3 and more jumps events are totally absent in the conditions we consider.}
\label{FigS1}
\end{figure}

\section{Quantum master equation}

The master equation for the density matrix reads
\begin{equation}\label{rho2res}
\dot{\rho}=\frac{1}{i \hbar}[\rho,\hat{{\cal{H}}_{s}}] + \mathcal{L}^{(1,2)} \, ,
\end{equation}
where losses are accounted for through Liouvillian operators in the usual Lindblad form for the two resonators modes,
$\mathcal{L}^{(1,2)}= \sum_{i=1,2} {\kappa_{i}} [\hat{a}_{i}\rho \hat{a}_{i}^{\dag} - 0.5(\hat{a}_{i}^{\dag}\hat{a}_{i}\rho-\rho\hat{a}_{i}^{\dag}\hat{a}_{i})]$.
Further sources of loss, such as nonlinear absorption (e.g. related to  the imaginary part of $\chi^{(3)}$) or pure dephasing, could also be added to Eq. \ref{rho2res} (see, e.g. \onlinecite{Ferretti2012,Ferretti2013}), but we neglect them here for simplicity. Moreover, the model can be generalized to include input and output quantum channels \cite{Flayac2013}. 

Single-time evolution and steady state numerical results of Eq.~\ref{rho2res} can be straightforwardly performed, as in Refs.~\onlinecite{Liew2010,Bamba2011,Ferretti2013}.
Here, we were additionally able to confirm the Montecarlo results (cyan curve in Fig.~2g of the main text).
The un-normalized and normalized two-times second-order correlation functions of cavity 1 were computed as
\begin{eqnarray}
{G^{(2)}}\left( {{t},{t'}} \right) &=& {\rm{Tr}}\left[ {\hat a_1^\dag {{\hat a}_1}{U_{{t} \to {t'}}}\left( {\hat a_1^\dag {{\hat a}_1}\hat \rho \left( {{t'}} \right)} \right)} \right]\\
{g^{(2)}}\left( {{t},{t'}} \right) &=& \frac{{{G^{(2)}}\left( {{t},{t'}} \right)}}{{{\rm{Tr}}\left[ {\hat a_1^\dag {{\hat a}_1}\hat \rho \left( {{t}} \right)} \right]{\rm{Tr}}\left[ {\hat a_1^\dag {{\hat a}_1}\hat \rho \left( {{t'}} \right)} \right]}} \,  ,
\end{eqnarray}
where ${\hat{\cal{U}}}_{{t} \to {t'}}(\hat O)$ is the propagator of the operator $\hat O$ from $t$ to $t'$ associated with Eq. \ref{rho2res}. The photon statistics produced within a time window $\Delta t=t_2-t_1$ is obtained from
\begin{equation}
{g^{(2)}}\left( {\Delta t} \right) = \frac{{\iint_{\Delta t} {{G^{(2)}}\left( {t,t'} \right) dt dt'}}}{{\iint_{\Delta t} {n_1( t )n_1( {t'} ) dt dt'}}}
\end{equation}
where $n_1( t )={\rm{Tr}}[ {\hat a_1^\dag {{\hat a}_1}\hat \rho \left( t \right)}]$. This exact calculation perfectly reproduces the Montecarlo wave function results within the error envelope, as it is reported in Fig.~2g (cyan curve).

\section{Photonic crystal cavities design}

Photonic crystal cavities allow to achieve record figures of merit today, such as ultra-small mode volumes and ultra-high quality factors~\cite{Notomi2010}.
One of the most used photonic crystal cavity designs is realized by removing three air holes in a triangular lattice~\cite{Akahane2003}, which is usually defined a L3 point defect.
Recently, a combination of fast simulation tools~\cite{Andreani2006} and genetic optimization~\cite{Minkov2014} have allowed to show that Q factors largely exceeding $10^6$ can be designed for such cavities, which was shown experimentally~\cite{Lai2014}.

For the photonic crystal cavities design used in this work, we started from a standard SOI photonic crystal membrane, with the silicon layer thickness of 220 nm. We set the lattice constant to $a=400$ nm and the holes radius to $r=112$ nm ($r/a=0.28$) to tune the cavity mode resonant wavelength in the relevant telecom band, i.e. $\lambda = 1.5$ $\mu$m ($\sim 0.825$ eV). The three holes along the cavity axis have been shifted by $s_1=120$ nm ($s_1 /a=0.3$) $s_2=100$ nm ($s_2 /a=0.25$), and $s_3=40$ nm ($s_3 /a=0.1$), to reach a theoretical (unloaded) $Q\sim 1.25 \times 10^6$. Since we aim at coupling these cavities with access and output waveguides, we thus allow the loaded Q-factor to be in the $0.8\times 10^6$ range.

A plot of the normalized electric field intensity, i.e. .the function $|\vec{\alpha} (\mathbf{r})|^2$ with $\int |\vec{\alpha}(\mathbf{r})|^2 \mathrm{d}\mathbf{r} =1$, is shown in Fig.~S2 for our optimized L3 cavity design, which was the building block for the the photonic crystal molecules in Fig.~3 of the main text.

\begin{figure}[t]
\includegraphics[width=0.45\textwidth,clip]{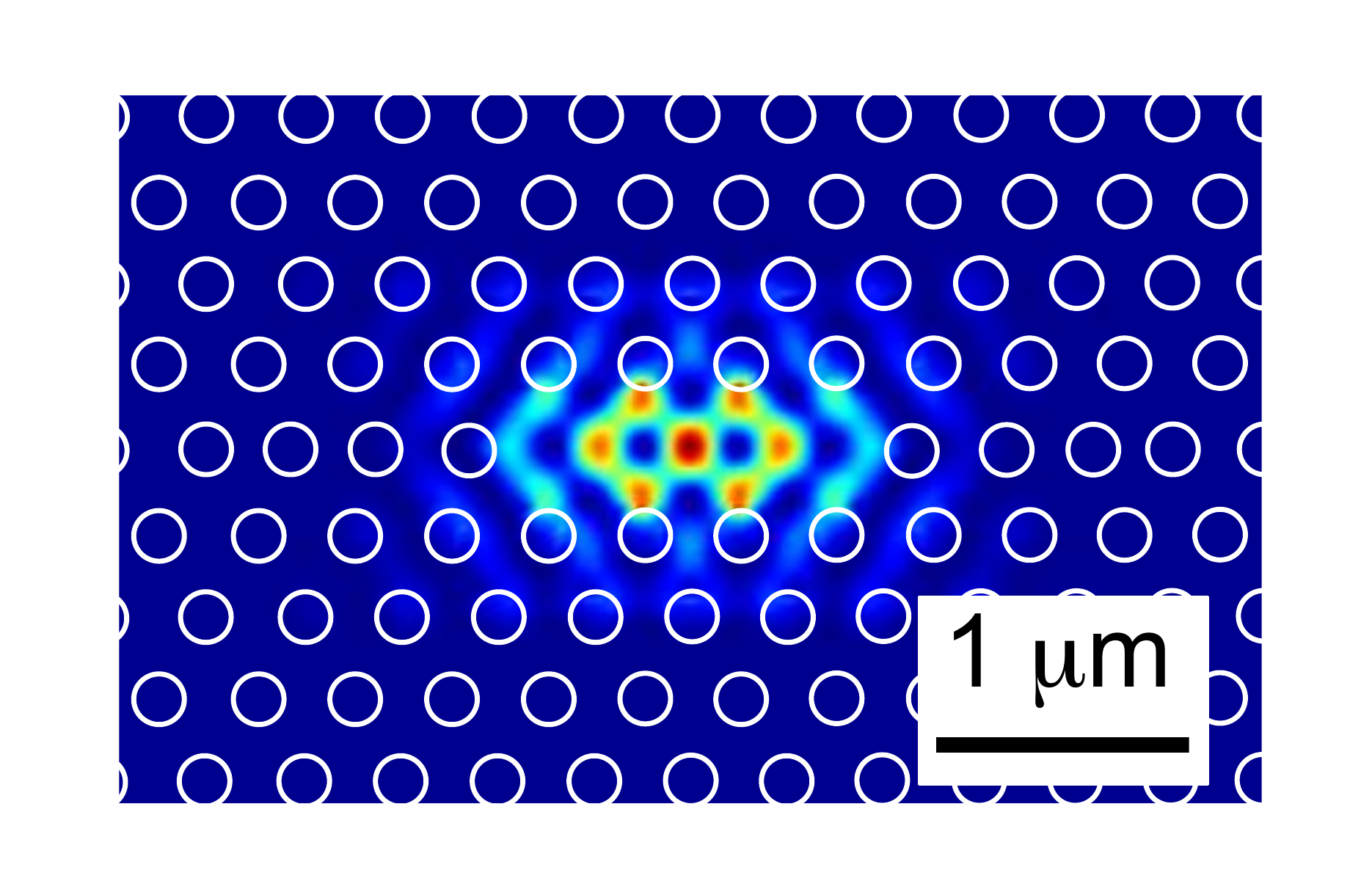}
\caption{Electric field intensity profile at the center of the silicon photonic crystal membrane.}
\label{FigS2}
\end{figure}

\section{Estimating the effective photon-photon interaction}

In this work, we focus on photonic nanostructures in silicon, which is a strongly nonlinear material already at the level of classical electromagnetic response. In particular, bulk silicon is characterized by a relatively large $\chi^{(3)}$ susceptibility, while nominally $\chi^{(2)} = 0$ (neglecting surface contributions) owing to the centrosymmetric nature of the elementary crystalline cell \cite{Boyd2008}. Strongly enhanced nonlinear effects have been already reported in L3 photonic crystal cavities~\cite{Galli2010}.

The photon-photon interaction energy in each resonator is given in terms of the material $\chi^{(3)}$ by the simplified expression \cite{Ferretti2012}
\begin{equation}\label{ham_kerr}
U= \frac{D(\hbar\omega_i)^2 } {8\varepsilon_0}
\int \mathrm{d}\mathbf{r} \,\frac{\chi^{(3)}(\mathbf{r})}{\varepsilon^2(\mathbf{r})} \, |{\alpha} (\mathbf{r}) |^4   \, ,
\end{equation}
where $\vec{\alpha} (\mathbf{r})$ is the three-dimensional cavity field profile, normalized as $\int |\vec{\alpha}(\mathbf{r})|^2 \mathrm{d}\mathbf{r} =1$, and $D$ represents the multiple contributions of the same order of magnitude given by the different elements of the $\chi^{(3)}$ tensor \cite{Boyd2008}.

From the calculated mode profile shown in Fig.~S2, the effective nonlinearity of such a silicon nanocavity can be estimated through Eq.~\ref{ham_kerr}, by using $\chi^{(3)}\sim 0.9\times 10^{-18}$ m$^2/$V$^2$, which is an appropriate order of magnitude for the elements of the bulk silicon third-order susceptibility tensor \cite{Boyd2008}, and D=24 \cite{Sipe1987}. For the cavity mode profile of our optimized photonics crystal structure, see Fig.~S2, a quantitative estimate of this integral results in $U\simeq 0.8\times 10^{-3}$ $\mu$eV, close to what was  assumed in the model calculations of the main text and confirming the order of magnitude estimates already given in the literature~\cite{Ferretti2013}.


\end{document}